\documentclass[english]{lni}
\IfFileExists{latin1.sty}{\usepackage{latin1}}{\usepackage{isolatin1}}

\usepackage{setspace}
\let\OLDthebibliography\thebibliography
\renewcommand\thebibliography[1]{
  \OLDthebibliography{#1}
  \setlength{\parskip}{1pt}
  \setlength{\itemsep}{1pt plus 0.3ex}
}

\usepackage{wrapfig}

\usepackage{titlesec}
\titlespacing*{\section}
{0pt}{2ex plus 1ex minus .2ex}{0.5ex plus .2ex}
\titlespacing*{\subsection}
{0pt}{1.5ex plus 1ex minus .2ex}{0.4ex plus .2ex}

\usepackage{enumitem}
\setlist[itemize]{itemsep=1pt, topsep=-3pt}
\setlist[enumerate]{itemsep=1pt, topsep=-3pt}

\usepackage{graphicx}
\usepackage{fancyhdr}
\usepackage{listings} %if lstlistings is used
\usepackage{changepage} %for changing topmargin on first page
\usepackage[figurename=Fig., tablename=Tab., small]{caption}[2008/04/01]
\usepackage{amsmath,amssymb,amsfonts}
\usepackage[ruled,vlined]{algorithm2e}
\usepackage{textcomp}
\usepackage{xcolor}
\usepackage{caption}
\usepackage{subcaption}
\usepackage{booktabs}
\usepackage{balance}
\usepackage{multirow}
\usepackage{url}
\renewcommand{\headrulewidth}{0.4pt} %horizontal line below header
\author{Cori Tymoszek\footnote{Michigan State University, East Lansing MI 48824} , \ Sunpreet S. Arora\footnote{Visa Research, Palo Alto CA 94306, sunarora@visa.com} , \ Kim Wagner\footnote{Visa Research, Palo Alto CA 94306, kwagner@visa.com} , and Anil K. Jain\footnote{Michigan State University, East Lansing MI 48824, jain@cse.msu.edu} }
\title{DashCam Pay: A System for In-vehicle Payments Using Face and Voice}
\begin{document}

\maketitle
%\vspace{-40pt}
\renewcommand{\refname}{References}
\setcounter{footnote}{4} %Change to the number of authors for a correct numbering of the foot notes
%\thispagestyle{titlepage}
%header setting after the second page
\pagestyle{fancy}
\fancyhead{} % clears header settings
\fancyhead[RO]{\small DashCam Pay: A System for In-vehicle Payments using Face and Voice \hspace{5pt} \thepage \hspace{0.05cm}}
\fancyhead[LE]{\hspace{0.05cm}\small \thepage \hspace{5pt} Cori Tymoszek, Sunpreet S. Arora, Kim Wagner and Anil K. Jain}
\fancyfoot{} % clears all footer settings
\renewcommand{\headrulewidth}{0.4pt} %line below header

\begin{abstract}
We present our ongoing work on developing a system, called \textit{DashCam Pay}, that enables in-vehicle payments in a seamless and secure manner using face and voice biometrics. A plug-and-play device (dashcam) mounted in the vehicle is used to capture face images and voice commands of passengers. Privacy-preserving biometric comparison techniques are used to compare the biometric data captured by the dashcam with the biometric data enrolled on the users' mobile devices over a wireless interface (e.g., Bluetooth or Wi-Fi Direct) to determine the payer. Once the payer is identified, payment is conducted using the enrolled payment credential on the mobile device of the payer. We conduct preliminary analysis on data collected using a commercially available dashcam to show the feasibility of building the proposed system. A prototype of the proposed system is also developed in Android. DashCam Pay can be integrated as a software solution by dashcam or vehicle manufacturers to enable open loop in-vehicle payments.
\end{abstract}
\begin{keywords}
Multi-biometrics, Privacy-preserving biometrics, In-vehicle payments
\end{keywords}

\section{Introduction}
\begin{wrapfigure}{r}{0.55\textwidth}
    \centering
    \includegraphics[width=0.5\textwidth]{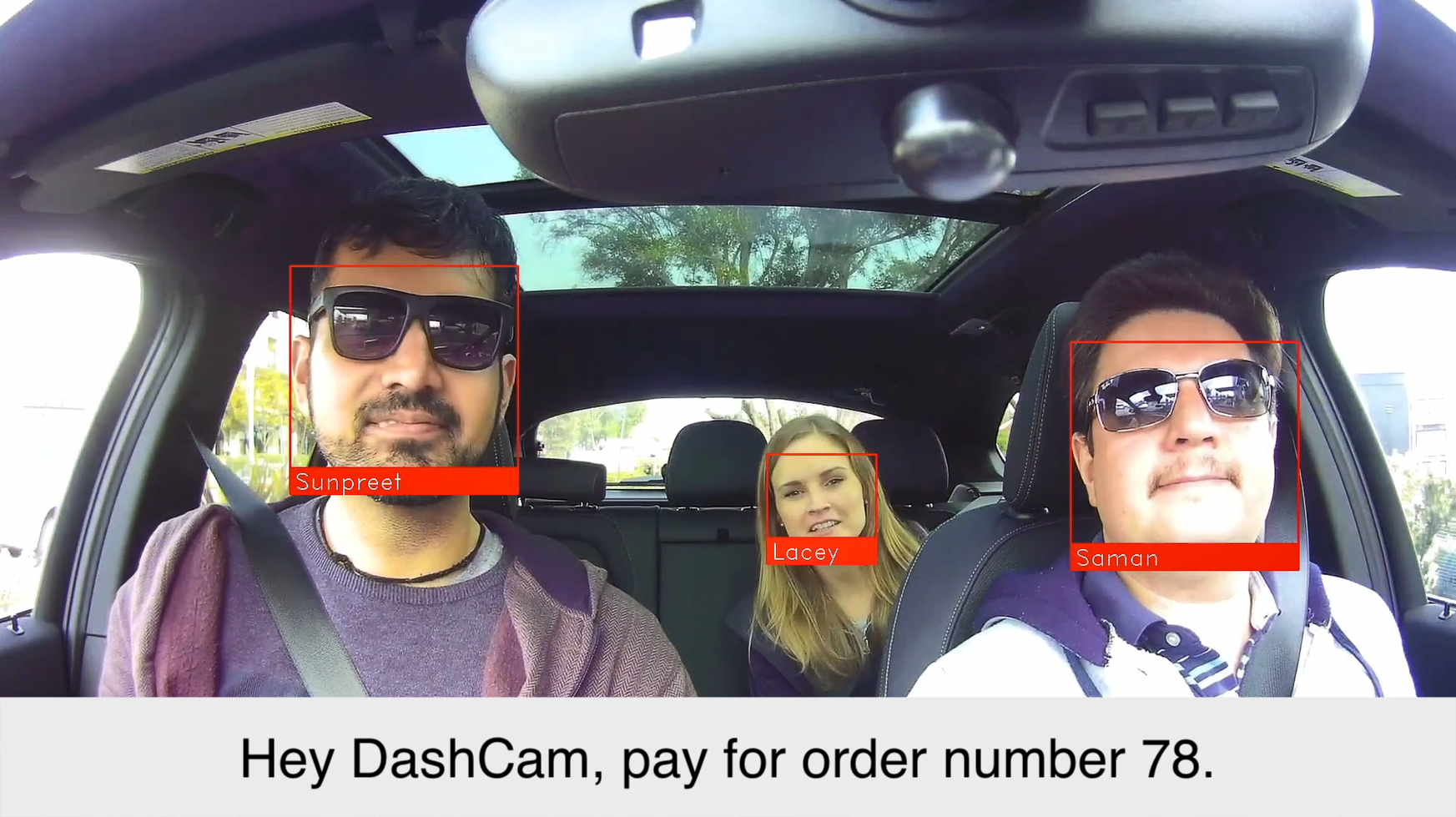}
    \caption{Users' faces and voices being recognized in real-time on data collected from a commercial dashcam.}
    \label{fig:example}
\end{wrapfigure}
The payments industry has witnessed the introduction of several innovative payment solutions in the last decade. Examples include contactless payments \cite{visa}, smartphone and smartwatch payment applications (e.g, Apple Pay \cite{apple-pay}), and mobile terminals (e.g., Square \cite{square}). A relatively new focus area for the payments industry, however, is in-vehicle payments where the goal is to enable the driver (and/or passengers) to pay in a seamless and secure manner. To this end, several vehicle manufacturers have integrated digital assistants, such as Siri and Alexa, in infotainment systems that can be used for making voice-based in-vehicle payments \cite{rain} \cite{amazon-ces}. In addition, some newer vehicles are “connected” and have integrated payment capabilities in the head-unit \cite{shell-gm}. Currently, such systems are \textit{closed-loop} and typically only enable payments for the owner of the vehicle, and not for any passenger riding in the vehicle in an \textit{open loop} manner. Also, these systems do not enable owners of older vehicles that lack these advanced capabilities to conduct in-vehicle payments. Alternatively, smartphone apps can be used for in-vehicle payments. While they are a viable option for passengers, smartphone usage can be quite distracting and is not recommended for drivers.

Given the limitations of existing in-vehicle payment solutions, we design an open loop system that uses face and voice biometrics to enable in-vehicle payments in a secure and seamless manner. The users of the system enroll their face and voice templates and a payment credential once in a mobile app, and pair their mobile device with a plug-and-play device (dashcam) mounted in a vehicle. To initiate payment, a passenger invokes the dashcam with a trigger phrase, for example, “Hey DashCam” and then issues a command, such as “Pay for gas at pump 5”. On hearing the trigger phrase, the dashcam takes a picture of the passengers in the vehicle using the in-cabin camera, as well as an audio recording of the command. The dashcam then initiates a privacy-preserving biometric comparison protocol with each of the connected mobile devices of the passengers. Both face and voice biometrics are used to determine which passenger issued the command and wants to pay. If a unique payer is determined, payment is conducted using the enrolled payment credential on the payer's device.

We collected over 120 minutes of combined audio and video data from 20 different subjects in five different vehicles at two different sites using a commercially available dashcam (Fig. \ref{fig:example}), and conducted preliminary analysis to show the feasibility of the proposed system. We also developed an Android-based prototype of our system. DashCam Pay can be integrated by dashcam or vehicle manufacturers to enable open-loop in-vehicle payments.

The proposed system can be useful in a variety of payment scenarios:

\begin{itemize}

    \item \textit{Drive-throughs}: To avoid the inconvenience of taking out cash or card from their wallet and handing it to a merchant representative at drive-throughs, a customer can directly use DashCam Pay for making payments.
    
    \item \textit{Toll booths}: A customer can use DashCam Pay to pay for toll while approaching the toll booth. The license plate number can be registered at the time of enrollment, and the location of the dashcam or the customer device can be used to determine the authority to which the payment is sent.
    
    \item \textit{Parking}: With DashCam Pay, a customer can pay for parking at a parking lot using their license plate number and/or space number. This identifier can be used along with customer device's location to issue payment to the relevant parking authority.
    
    \item \textit{Gas stations}: Instead of inserting payment card at a gas station or using cash, a customer can use DashCam Pay to pay for gas using the gas station number. The merchant can be determined by the customer device's location.
    
    \item \textit{Retail stores}: The proposed system can be deployed in similar spirit on merchant terminals. This would allow customers to make in-store payments without interacting with their mobile devices or using payment cards.
\end{itemize}

%\subsection{Key Contributions}
The major contributions of this work are:
\begin{enumerate}
    \item Design of a multi-biometric system based on face and voice that enables secure and seamless in-vehicle payments; the fundamental system design is reusable for other payment scenarios such as in-store payments.
    
    \item A privacy-preserving biometric comparison protocol between a device mounted in a vehicle and mobile devices of passengers in the vehicle over a wireless interface to identify the payer.
        
    \item Preliminary analysis to show the feasibility of building the proposed system for in-vehicle payer identification.
    
    \item Development of an Android prototype of the system.

\end{enumerate}

\section{System Design}
In this section, we describe the objectives of the proposed system, the system modules as well as the privacy-preserving biometric comparison protocol used in the system.

\subsection{Objectives and Design Choices}
Broadly, the three key objectives were to build an in-vehicle payment system that is (1) easy to use, (2) secure, and (3) operates in real-time. Below we highlight the key design choices in light of these goals.

\begin{itemize}
    \item \textit{Convenience}: Driven entirely by voice in a hands-free manner. In addition, two different biometric modalities, face and voice are used to reduce false matches in practice. Face is used to determine candidate payers from passengers in a vehicle, and voice is used to identify the payer from the candidate payers at the time of payment.
    \item \textit{Security}: To ensure the dashcam does not learn anything about a user's enrolled biometric data since the dashcam may not be owned by the user, the enrolled data is compared with biometric data captured for authentication by the dashcam in the encrypted domain. This protocol is described in detail in Section \ref{sec:ppbc}.
    \item \textit{Real-time operation}: Face comparisons are conducted in advance at the time a user's mobile device is connected to the dashcam, and periodically throughout the ride to determine candidate payers, and only voice comparisons are conducted at the time of payment to identify the payer. Fusion of face and voice in this manner supports real-time operation of the system.
\end{itemize}

\begin{figure}[t]
        \centering
        \includegraphics[trim=105 70 105 70, clip,width=0.65\columnwidth]{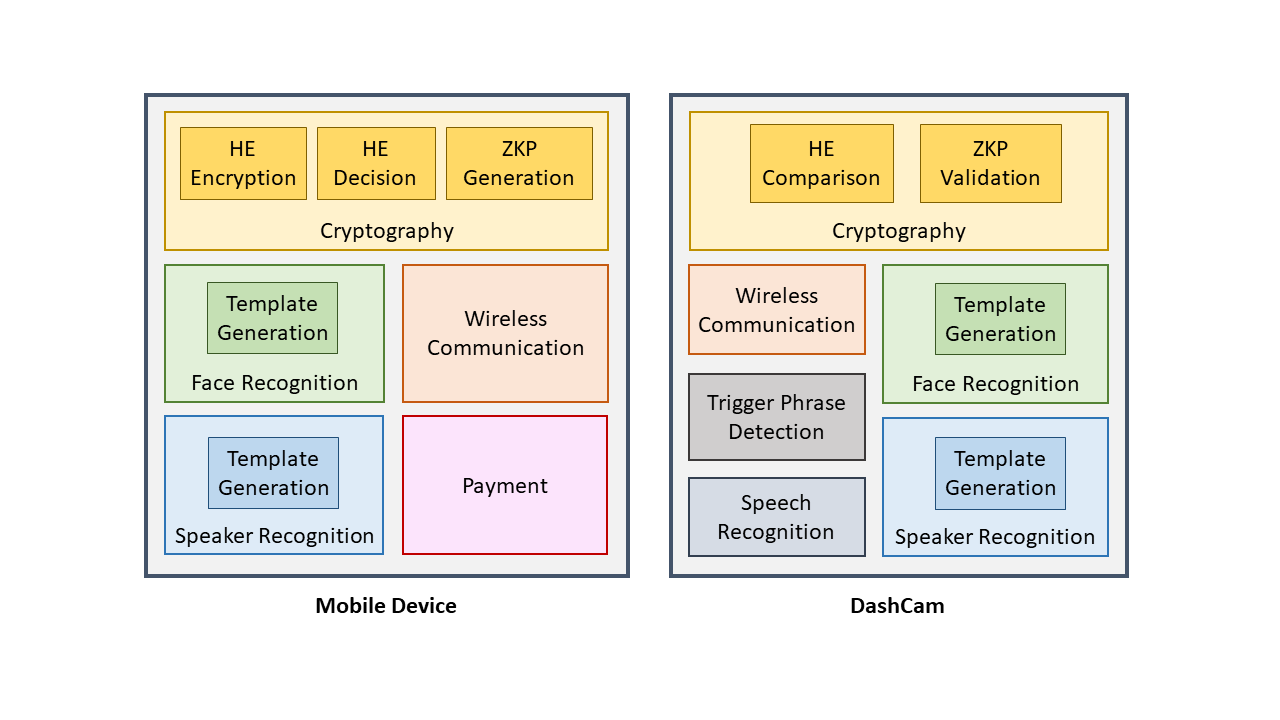}
    \caption{Component diagram showing the main system modules on the mobile device and dashcam.}
    \label{fig:dashcam_component_diag}
\end{figure}

\subsection{System Modules}
Given the aforementioned objectives, the system design includes the following modules (Fig. \ref{fig:dashcam_component_diag}):

\begin{itemize}
    \item \textit{Wireless communication}: To connect users' personal devices and the dashcam over a wireless channel such as BLE or Wi-Fi Direct
    \item \textit{Trigger phrase detection}: To invoke the dashcam in a hands-free manner by speaking a specific phrase (e.g., "Hey DashCam...").
    \item \textit{Face recognition}: To enroll a user's face on the user's personal mobile device, and identify candidate payers.
    \item \textit{Speech recognition}: For parsing and execution of a user's voice commands.
    \item \textit{Speaker recognition}: To enroll a user's voice on the user's personal mobile device, and identify the payer.
   \item \textit{Cryptography}: To enable privacy-preserving biometric comparison as well as ensure secure data exchange between user's mobile devices and dashcam. See Section \ref{sec:ppbc} for more details.
   \item \textit{Payment}: To enable user devices to enroll a payment credential, e.g., a card or wallet and conduct payments with the enrolled credential.
\end{itemize}

\subsection{Privacy-Preserving Biometric Comparison}
\label{sec:ppbc}

\begin{wrapfigure}{r}{0.6\textwidth}
    \centering
       \includegraphics[trim=80 60 60 50, clip,width=0.65\textwidth]{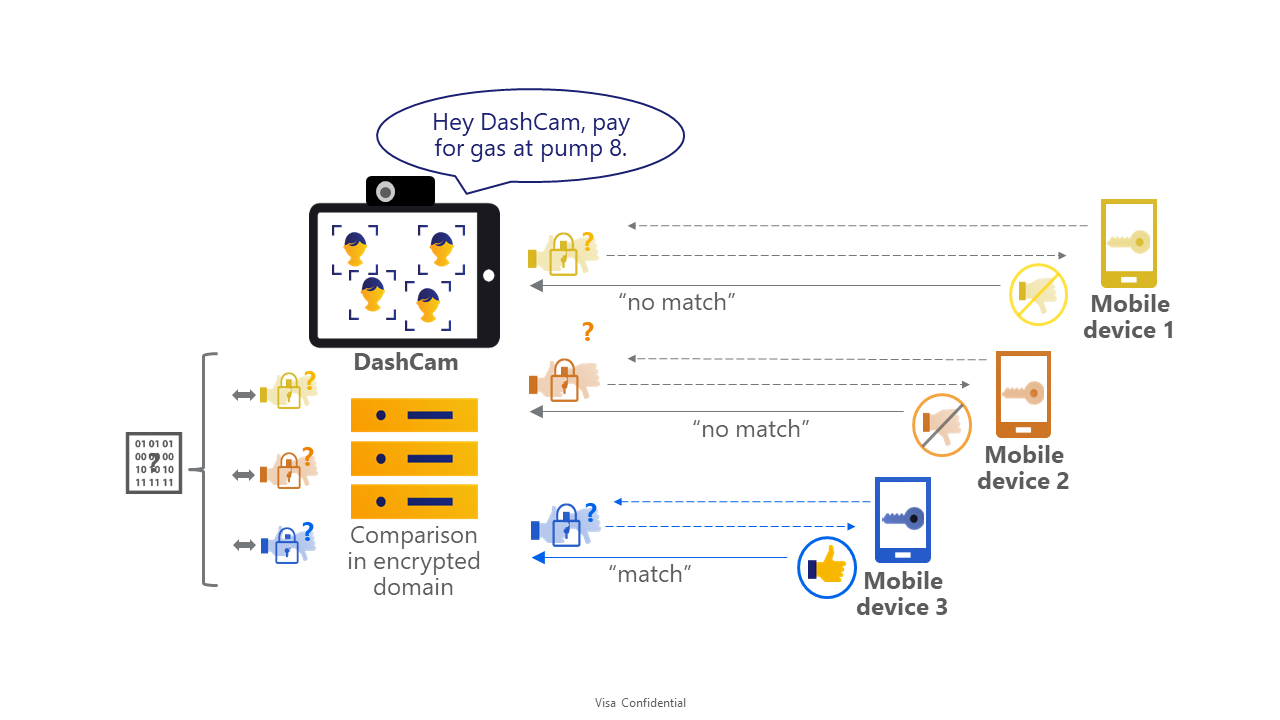}

    \caption{Privacy-preserving biometric comparison protocol between dashcam and mobile devices of users connected to dashcam} 
    \label{fig:protocol}
\end{wrapfigure}
Several privacy regulations in different geographies, e.g., European Union (EU) General Data Protection Regulation(GDPR) 2016/679 \cite{gdpr}, categorize biometric data as sensitive information and suggest restricting access to such data under the \textit{right to privacy}. In the proposed system, the dashcam may not be owned by a user, and may be installed in a vehicle that is not owned by the user. In order to maintain the privacy of sensitive biometric data, it is important to ensure that the dashcam only learns that a biometric match was obtained without learning anything more about the enrolled biometric data of a user. To this end, the proposed privacy-preserving biometric comparison protocol (Fig. \ref{fig:protocol}) leverages fully homomorphic encryption techniques (e.g., see \cite{Naresh_Boddeti_2018}). Compared to other biometric template security methods, such as biometric cryptosystems or cancelable templates \cite{rathgeb2011survey}, state-of-the-art fully homomorphic encryption-based techniques not only satisfy \textit{irreversibility}, \textit{unlinkability} and \textit{renewability} - the three security requirements stated in ISO/IEC IS24745:2011 \cite{iso-24745}, they do not degrade biometric system recognition performance.

Let $D$ represent a dashcam and the set $M = \{M_1, M_2, ... M_n\}$ represent the $n$ mobile devices of enrolled users. Following is the sequence of steps followed at the time of (i) enrollment on a user's mobile device, (ii) connection of a user's device to dashcam, and (iii) identification of candidate payers using face, or determination of payer from candidate payers using voice.

\begin{enumerate}
    \item \textit{Enrollment}: When user $i$ enrolls on device $M_i$:
    
    \begin{enumerate}
        \item Enrollment template $ET_i$ of user $i$ is generated
         \item A key generation function is used to generate a homomorphic public-private key pair $\{K^{priv}_i, K^{pub}_i\}$ on $M_i$.
        \item $ET_i$ is encrypted using $K^{pub}_i$ to generate $Enc(ET_i)$.
    \end{enumerate}
   
    \item \textit{Connection}: When $M_i$ connects with $D$ (over a wireless interface such as BLE, WiFi):
    
    \begin{enumerate}
        \item A device ID $m_i$ is generated for device $M_i$.
        \item $Enc(ET_i)$ is sent to $D$ along with $K^{pub}_i$ and $m_i$.
    \end{enumerate}
    
    \item \textit{Payer identification}: Let $Enc(ET)$ be the set of biometric templates and $K^{pub}$ be the set of corresponding public keys of connected users accumulated at $D$ at the time of payer identification.
    
    \begin{enumerate}
        \item Biometric template $AT_i$ is generated from biometric data captured for passenger $i$ on $D$.
        \item  The set of generated biometric templates $AT$ is compared with enrollment templates $Enc(ET)$ on $D$ using homomorphic encryption, each with the corresponding public key $K^{pub}_i$ to compute a set of encrypted similarity scores $Enc(S)$.
        \item Each encrypted score $Enc(S_i)$ is sent to corresponding device with ID $m_i$ where it is decrypted using the private key $K^{priv}_i$ to yield $S(i)$.
        \item $m_i$ compares $S(i)$ to a predefined similarity score threshold $t$ (based on estimated false accept rate) to determine if $S(i)>t$ i.e. a match was obtained.
        \item $m_i$ generates a zero-knowledge proof $Z_i$ that $Enc(S_i)$ was correctly decrypted, and to prove that a match or a non-match is obtained. $m_i$ sends $Z_i$ to $D$.
        \item $D$ validates $Z_i$ to ascertain if a match is obtained. Use of zero-knowledge proof \cite{zeroknowledge} ensures that $D$ only learns whether a match is obtained from device ID $m_i$, and does not learn anything else about user $i$'s enrolled biometric data on $m_i$.
        \item Let $m_s$ be the unique device ID for which match is obtained. $D$ sends $m_s$ requisite details (e.g. order number) to conduct payment.
    \end{enumerate}
\end{enumerate}

In case the protocol fails to yield a unique match or multiple matches are obtained, a user recourse action is taken. Examples of recourse actions include requesting the user to try again or use an alternative method to pay.

\section{Practical Challenges}
In-vehicle payer identification using face and voice is a challenging problem. Below we discuss the key challenges.

\begin{itemize}
\item \textbf{Facial pose} changes as a normal consequence of driving in the vehicle for both the driver and the passengers. An example of this difficulty with pose is shown in Fig. \ref{fig:challa}. However, it is reasonable to assume that a user would cooperate when they want to pay using DashCam Pay. Extreme facial pose changes are, therefore, unlikely to be encountered in practice. Also, state of-the-art face recognition algorithms are robust to reasonable pose changes \cite{pifr-ding2015}\cite{pifr-tran2017}.

\begin{figure}[t]
\centering
    \begin{subfigure}[b]{0.55\columnwidth}
    \centering
    \includegraphics[width=\columnwidth]{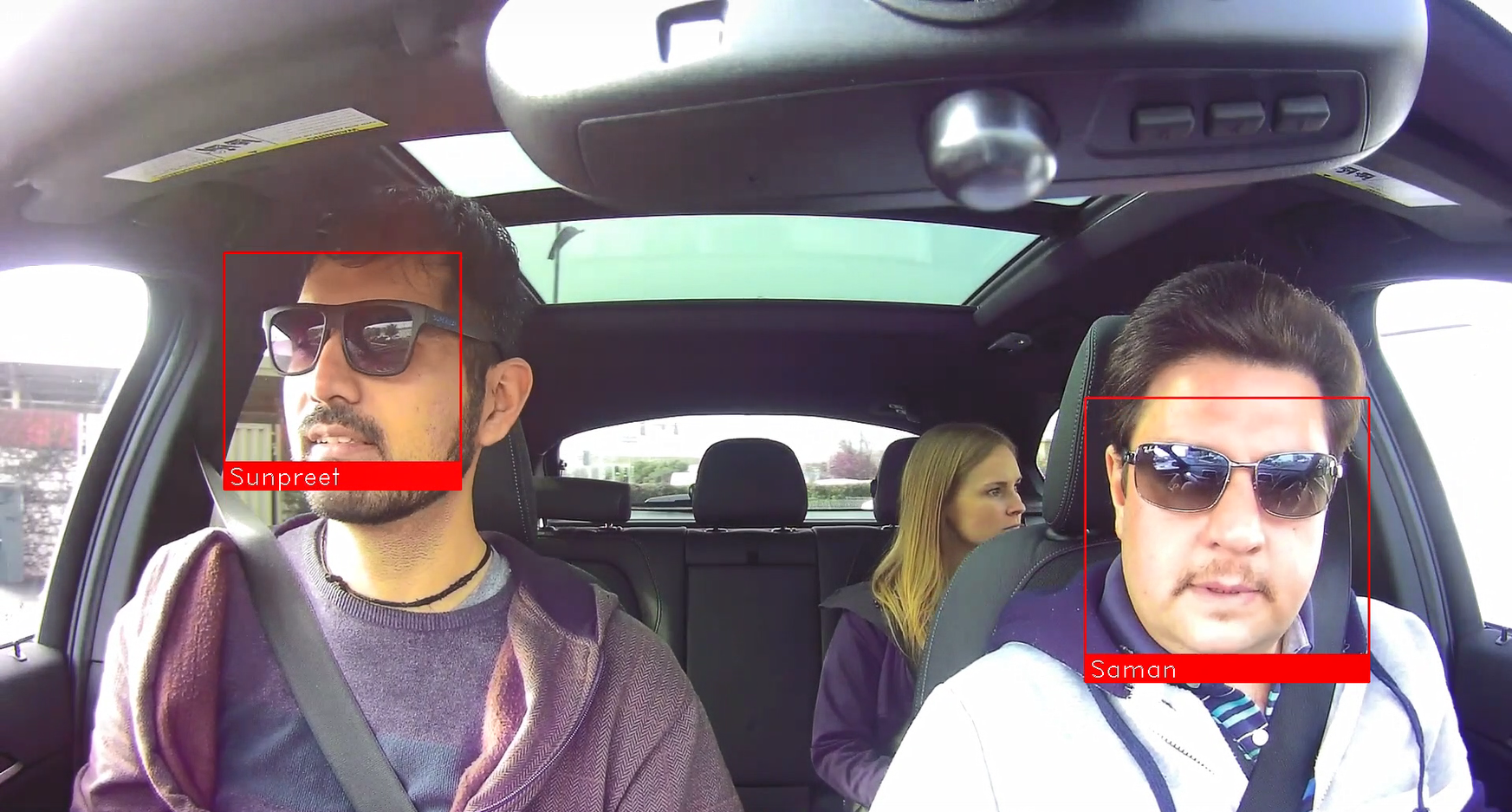}
    \caption{}
    \label{fig:challa}
    \end{subfigure}
    \begin{subfigure}[b]{0.55\columnwidth}
    \centering
    \includegraphics[width=\columnwidth]{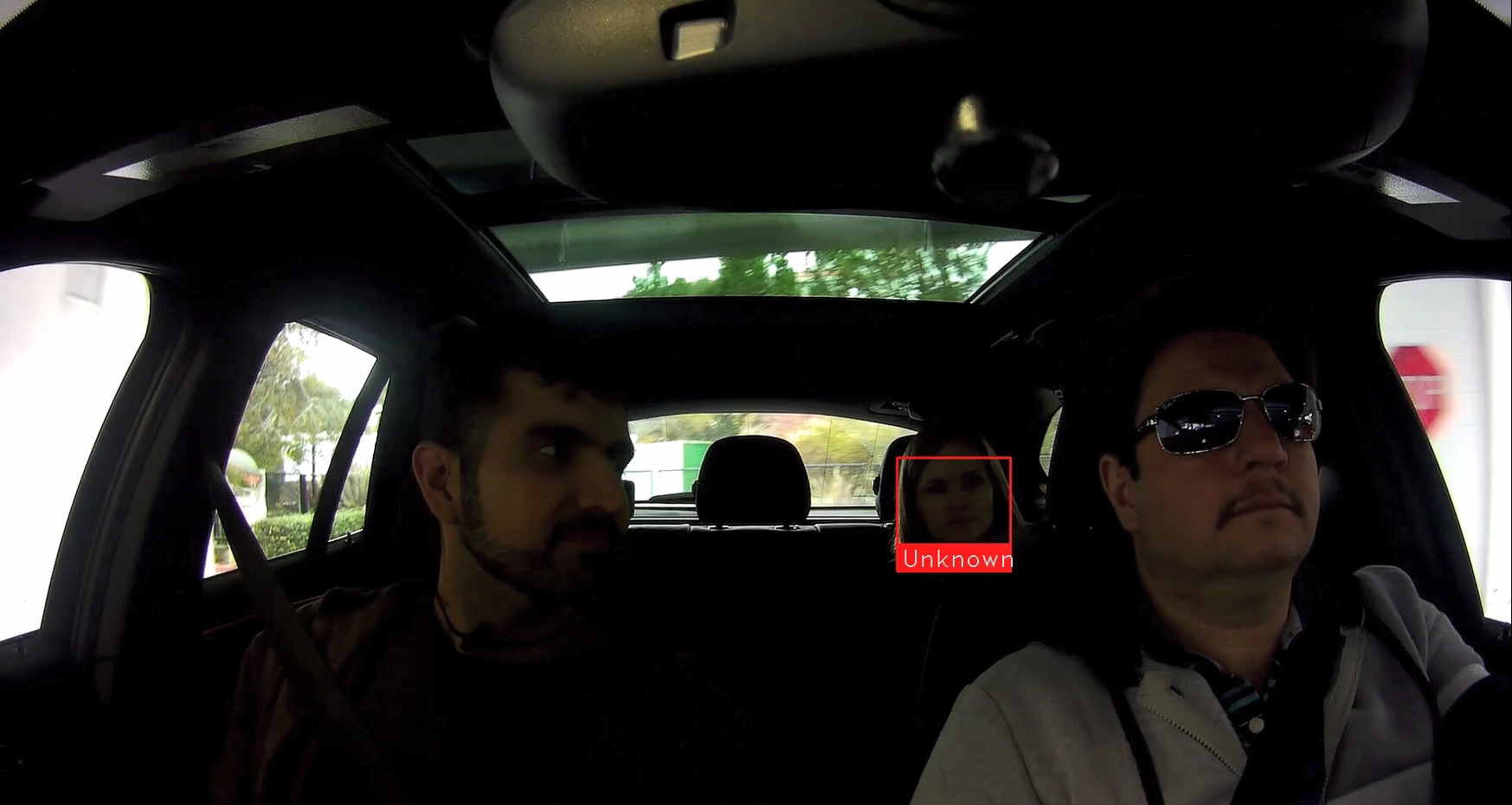}
    \caption{}
    \label{fig:challb}
    \end{subfigure}
    \begin{subfigure}[b]{0.55\columnwidth}
    \centering
    \includegraphics[width=\columnwidth]{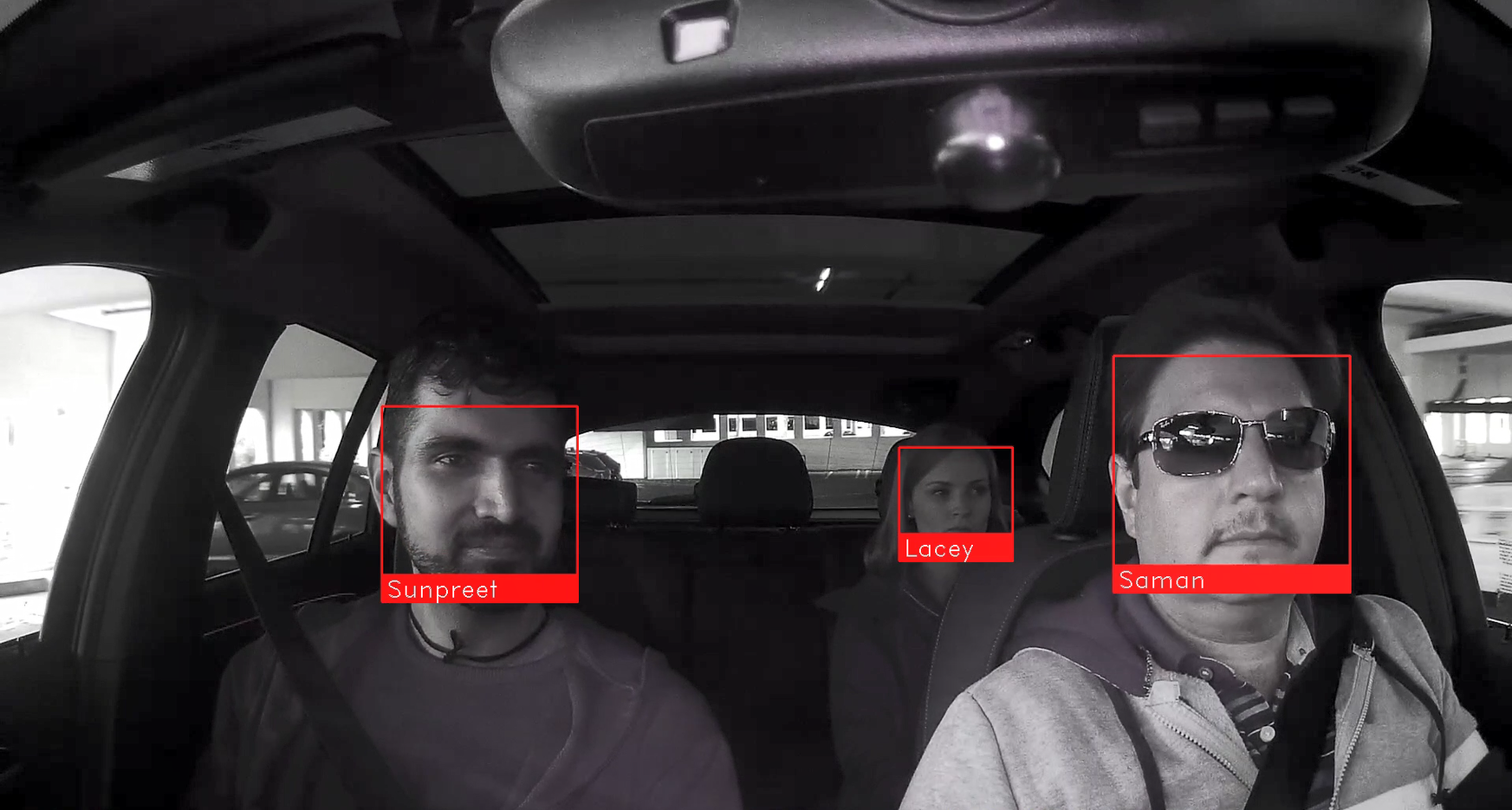}
    \caption{}
    \label{fig:challc}
    \end{subfigure}
  \caption{In-vehicle face recognition challenges: (a) Face is not detected for passenger in the backseat when pose is extreme, and (b) Visible spectrum image in low lighting shows failure to detect faces for front-seat passengers, and failure to recognize backseat passenger. (c) Infrared image in low lighting shows all three passengers are successfully detected and recognized.}
  \label{fig:chall}
\end{figure}

\item \textbf{Lighting} variations in the vehicle's cabin. For example, the vehicle may pass through a tunnel or garage, under foliage, or near brightly lit signs while a passenger is interacting with the dashcam. A dashcam with infrared capture capabilities in low lighting environments can be used to mitigate this problem to a large extent. Figs. \ref{fig:challb} and \ref{fig:challc} show poor performance in the visible spectrum, and good performance in the infrared spectrum, in the low lighting environment respectively.

\item \textbf{Matching of heterogeneous images} as commercial dashcams capture face images of passengers typically with a wide-angle lens and in infrared mode in low-light situations. On the other hand, the enrolled face images of a user are taken on a mobile device in the visible spectrum. Fortunately, the face recognition community has addressed many of these challenges\cite{stanli-2007}\cite{ranhe-hfr2018}\cite{klare-2012}.

\item \textbf{Poor quality speech recording} due to background noise. There may be a radio playing in the vehicle, nearby construction, sirens, or general road noise from the vehicle. While it is reasonable to assume that the driver would reduce the volume of the radio before using DashCam Pay, other noise levels typically cannot be predicted or controlled. Note that speech and speaker recognition community has developed models that are robust to such background noise \cite{nrsr-seltzer2013}\cite{nrsr-enhance2015}.
\end{itemize}

\section{Feasibility Study}\label{ref:feasibility}
To assess the feasibility of performing biometric recognition in real-life, in-vehicle scenario, we collected data using a commercial dashcam device (Vantrue N2 Pro Uber Dual Dash Cam\footnote{Uses Sony Exmor IMX323 sensor with four infrared LEDs for dual 1920x1080p visible and infrared spectrum video.}).

\subsection{Data}

\begin{wraptable}{r}{0.6\textwidth}
\centering
\resizebox{0.55\textwidth}{!}{%
\begin{tabular}{l|l|l|l|l}
\toprule
\textbf{\#Subjects} & \textbf{\#Vehicles} & \textbf{ Visible rec.} & \textbf{Infrared rec.} & \textbf{Audio rec.}\\
\midrule
20 & 5 & 80 mins. & 40 mins. & 40 mins.\\
 \bottomrule
\end{tabular}
}%
\caption{Summary of the data collected using a commercial dashcam for feasibility experiments.}
\label{tab:data}
\end{wraptable}
The data consists of over 120 minutes of combined audio and video streams from 20 subjects in five different vehicles at two different sites, captured in both visible and infrared spectrum (see Tab. \ref{tab:data}). An example of a captured video frame is shown in Fig. \ref{fig:example}. The dashcam's default settings are used, whereby it automatically switches to infrared wavelength video recording when low lighting levels are detected. For each subject, approximately four minutes of visible spectrum and two minutes of infrared spectrum video are captured. Audio data is captured simultaneously, with scripted verbal commands comprising approximately two minutes of total audio per subject.

Each subject is recorded while giving a number of voice commands for four different use cases: fuel, toll, parking and fast food. These commands are constructed specifically for ease of use, and to ensure that the user includes the necessary information for each proposed use case (e.g., parking space number, order number). Example sentences include: (i) "Hey DashCam, pay for parking at space number 5208.", (ii) "Hey DashCam, pay for order number 120.", (iii) "Hey DashCam, pay for toll.", and (iv) "Hey DashCam, pay for gas at pump six."

\begin{table}[t]
\centering
\resizebox{0.6\columnwidth}{!}{%
\begin{tabular}{l|l|l}
\toprule
\textbf{Experiment} & \textbf{Software} & \textbf{Performance} \\ \midrule
\multirow{2}{*}{Face Detection} & \multirow{2}{*}{MTCNN \cite{zhang2016joint}} & TPR=99.1\%,\\
&&FPR=0.01\%\\
\midrule
\multirow{2}{*}{Face Recognition (1:N)} & \multirow{2}{*}{FaceNet \cite{facenet}} & TPR=98.9\%\\
&&@FPR=0.01\%\\
\midrule
\multirow{2}{*}{Trigger Phrase Detection} & \multirow{2}{*}{Mycroft AI Precise \cite{mycroft}} & TPR=98.2\%,\ \\
&&FPR=1\%\\
\midrule
\multirow{2}{*}{Speech Recognition} & \multirow{2}{*}{DeepSpeech \cite{deepspeech}} & \multirow{2}{*}{WER=3.65\%}\\&&\\
\midrule
\multirow{2}{*}{Speaker Recognition (1:N)} & \multirow{2}{*}{COTS} & TPR=98.4\%\\ &&@FPR=0.01\%\\
 \bottomrule
\end{tabular}%
}
\caption{Feasibility experiment results. TPR = True Positive Rate, FPR = False Positive Rate, WER = Word Error Rate (post dictionary-based correction). N is the number of subjects in the vehicle.}
\label{tab:feasibility}
\end{table}

\subsection{Evaluation}
Next, we evaluated one state-of-the-art biometric recognition algorithm for each module to determine the feasibility of building the proposed system. Face and trigger phrase detection performance is measured using true positive detection rate (TPR) at a fixed false positive detection rate (FPR). Small-scale identification experiments (1: N where N is the number of subjects in the vehicle) are performed to assess face and speaker recognition performance, and the recognition performance is measured using true positive identification rate (TPIR) at a fixed false positive identification rate (FPIR). Speech recognition performance is evaluated using Word error rate (WER).

\begin{itemize}
    \item \emph{Face Detection:} Pre-trained multi-task convolutional neural network (MTCNN)-based face detector \cite{zhang2016joint} is used for face detection experiments. Face locations of subjects in the vehicle are manually labelled with a bounding box in each frame of a recorded video. Face detection performance is computed in aggregate over all frames in a video. MTCNN-based method resulted in $99.1\%$ TPR at $0.01\%$ FPR on the collected data. Face detection failure is observed in captured frames with (i) extreme facial pose of subjects, or (ii) lack of proper illumination. Figs. \ref{fig:challa} and \ref{fig:challb} show failure examples. Note, however, that extreme facial pose is less likely to be encountered in practice at the time of recognition as the subjects cooperate while interacting with the dashcam.
    
    \item \emph{Face Recognition:} Pre-trained FaceNet model \cite{facenet} is used for face recognition experiments. An image of each subject in the vehicle captured using their smartphone along with their subject identifier is assumed to be enrolled in the gallery. Detected faces in each video frame are assumed to be probe images. Subject identifiers for probe images were manually labelled. FaceNet yielded TPIR of $98.9\%$ at FPIR of $0.01\%$. Fig. \ref{fig:challb} shows failure of face recognition in low lighting environment in the visible spectrum.
    
    \item \emph{Trigger Phrase Detection:} Mycroft AI precise \cite{mycroft} is used for trigger phrase detection experiments. Positive examples corresponding to training speech samples containing the trigger phrase, and negative examples referring to speech samples that do not contain the trigger phrase are manually created from the data. A custom detector is trained for the phrase "Hey DashCam". Audio data is split into and testing (50-50 split: 10 subjects for training and 10 for testing). For each positive sample, five negative samples are used in the training set to reduce false positives. The trained detector yielded TPR of $98.2\%$ at FPR of $1\%$ on the test data.
    
    \item \emph{Speech Recognition:} Mozilla's open-source DeepSpeech implementation \cite{deepspeech} is used for speech recognition experiments. Audio data extracted from the video streams is manually labelled based on the audio commands spoken by the subjects. The pre-trained model is first evaluated on the extracted audio data. Although the pre-trained model achieved $7.5\%$ WER on the LibriSpeech test-clean benchmark\cite{mozilla-dec2019}, $25.4\%$ WER is obtained on the collected data. The model frequently failed to recognize relevant words like ``dashcam", ``gas", and ``parking". Therefore, a portion of the collected audio data is used for fine-tuning the pre-trained model. Cross-validation experiments (20 percent for fine-tuning, and 80 percent for evaluation) are conducted to measure performance after fine-tuning. Post fine-tuning, average WER reduced to $8.7\%$. The majority of errors ($68\%$) are partial; for instance, "parking" recognized as "parting". Such errors were corrected using a dictionary of permitted words/phrases in a command. The following heuristic is used: if edit distance between the detected word and the closest word in the dictionary was less than or equal to 2, the detected word is auto-corrected to the closest word in the dictionary. Using this approach, overall WER reduced to $3.65\%$.
    
    \item \emph{Speaker Recognition:} A commercial off-the-shelf system (COTS) is used for speaker recognition experiments. Three samples of an audio command from each subject are enrolled in the system. TPIR of $98.4\%$ is obtained at FPIR of $0.01\%$.
  
\end{itemize}

Tab. \ref{tab:feasibility} summarizes the evaluation results. Overall results show the feasibility of conducting in-vehicle payer identification using face and voice biometrics.

\section{Prototype Development}
\begin{wrapfigure}{r}{0.4\textwidth}
    \centering
    \includegraphics[width=0.35\textwidth]{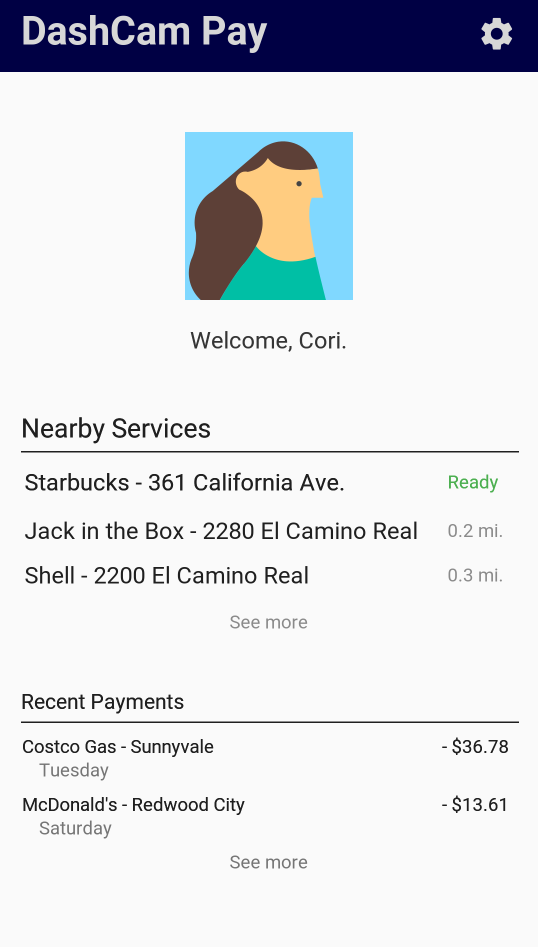}
    \caption{Dashboard of mobile device app, showing nearby merchants that accept DashCam Pay, as well as recent payments made by the user (simulated).}
    \label{fig:mockup}
    \end{wrapfigure}

Commercially available dashcams are not programmable, so we implemented the prototype as a software stack in Android, wherein one Android device is mounted in-cabin to act as the dashcam, and the others act as personal devices of the passengers (Fig. \ref{fig:mockup}).

\subsection{Implementation Details}
\begin{itemize}
    \item \textit{Wireless communication}: Google's Nearby Connections API\cite{nearbyconnect} is used to enable wireless communication between mobile devices and dashcam. This API uses both Bluetooth and Wi-Fi Direct to connect devices. The combination of modalities allows for the high speeds and versatility of Wi-Fi, with Bluetooth as a fallback. These technologies are also platform-agnostic, and can be implemented on any realization of the dashcam device, whether it be a mobile device, or a Raspberry Pi-like device (via Google Things)\cite{Antonioli2019NearbyTR}.

    \item \textit{Trigger phrase detection}: Android does not provide functionality for a third-party application (e.g., Mycroft AI precise \cite{mycroft}) to continuously access the device microphone. Hence, trigger phrase detection is implemented using Google Assistant as a bridge. This allows the user to give a command such as "Hey Google, open DashCam Pay" or "Hey Google, tell DashCam Pay to pay for gas at pump number four," to trigger the same authentication and payment flow described above.
    
    \item \textit {Face recognition}: FaceNet\cite{facenet} provided reasonable face recognition performance in our evaluation, and the model can be ported to the Android platform. Hence, it was used in our prototype. Other state-of-the-art models like SphereFace\cite{liu2017sphereface} and CosFace\cite{wang2018cosface} can also be used.
    
    \item \textit{Speech recognition}: The fine-tuned DeepSpeech model \cite{deepspeech} provided good performance in our evaluation. However, DeepSpeech's Android library did not support the use of custom models at the time of prototype development. So, Kaldi and Android's SpeechRecognizer class \cite{speechrecognizer} were considered as potential alternatives. However, no pre-built Android library is provided for Kaldi, and running the provided binaries on Android requires root access. Given this, we chose Android's SpeechRecognizer class for the prototype. This library is Android-specific, and provides state-of-the-art results~\cite{He_2019}.
    
    \item \textit{Speaker recognition}: The commercial off-the-shelf (COTS) speaker recognition system yielded exceptional results in our evaluation, and also provides support for Android platform. So, it was used in our prototype.
    
    \item \textit{Cryptography}: A commercial off-the-shelf (COTS) SDK is used to implement the privacy-preserving biometric comparison protocol described in Section \ref{sec:ppbc}. The SDK uses fully homomorphic encryption, and is able to perform generation of keys and encrypted biometric comparisons in real-time ($\leq$ 1 sec.). Similarity functions used for face and speaker recognition (cosine, euclidean) are encoded in the encrypted domain such that they do not impact the biometric comparison performance. The SDK also supports efficient generation and validation of zero-knowledge proofs.
    
    \item \textit{Payment}: The payment credential enrollment and usage is currently simulated in our prototype. We plan to integrate real-time payment capabilities in the next iteration of the prototype.
    
\end{itemize}

\subsection{Preliminary Evaluation}
We conducted preliminary evaluation of our prototype with 5 users. In 10 different trials, the users were able to invoke the dashcam device successfully. Furthermore, the dashcam device was able to discover the connected mobile devices of users, and exchange encrypted biometric data over both WiFi and BLE channels to successfully identify the payer. One potential issue is the increase in size of templates post encryption. Almost a 25 fold increase in the size of biometric templates is observed using the COTS SDK. However, since the enrollment data is sent to the dashcam at the time of connection, this is not a critical bottleneck for the operational performance of the prototype as long as there is sufficient time (about 10 seconds over BLE and 2 seconds for WiFi) for enrollment data transfer post connection before payment. We plan to source a more-efficient homomorphic encryption implementation to overcome this limitation in the next iteration of the prototype.

\section{Conclusions and Ongoing Work}
We present our ongoing work on developing a system for conducting in-vehicle payments in a secure and seamless manner using face and voice biometrics. The system uses privacy-preserving biometric comparison techniques to identify the payer from the passengers in a vehicle. We show the feasibility of the system on data collected from a commercial dashcam. An Android prototype of the system is also developed. Our system can be integrated by any vehicle or dashcam manufacturer to enable open-loop in-vehicle payments.
Once we finish refining our prototype, we plan to conduct comprehensive end-to-end testing and real-world pilots with merchants to refine the system before deployment in production.

\balance

\bibliographystyle{lnig}
\bibliography{lniguide}

\begin{thebibliography}{GMR85}

\bibitem[Am20]{amazon-ces}
Amazon announces new automotive products and solutions at CES 2020.
\newblock
  https://blog.aboutamazon.com/devices/new-automotive-products-and-solutions-at-ces-2020,
  2020.

\bibitem[Ap]{apple-pay}
Apple Pay.
\newblock https://www.apple.com/apple-pay/.

\bibitem[ATR19]{Antonioli2019NearbyTR}
Antonioli, Daniele; Tippenhauer, Nils~Ole; Rasmussen, Kasper~Bonne: Nearby
  Threats: Reversing, Analyzing, and Attacking {G}oogle's '{N}earby
  {C}onnections' on {A}ndroid.
\newblock In: NDSS.
\newblock 2019.

\bibitem[Bo18]{Naresh_Boddeti_2018}
Boddeti, Vishnu~Naresh: Secure Face Matching Using Fully Homomorphic
  Encryption.
\newblock 2018 IEEE 9th International Conference on Biometrics Theory,
  Applications and Systems (BTAS), Oct 2018.

\bibitem[DXT15]{pifr-ding2015}
{Ding}, C.; {Xu}, C.; {Tao}, D.: Multi-Task Pose-Invariant Face Recognition.
\newblock IEEE Transactions on Image Processing, 24(3):980--993, March 2015.

\bibitem[GMR85]{zeroknowledge}
Goldwasser, S; Micali, S; Rackoff, C: The Knowledge Complexity of Interactive
  Proof-Systems.
\newblock In: Proceedings of the Seventeenth Annual ACM Symposium on Theory of
  Computing.
\newblock STOC ’85, Association for Computing Machinery, New York, NY, USA,
  p. 291–304, 1985.

\bibitem[Go19]{nearbyconnect}
Google Developers {N}earby {C}onnections {API}.
\newblock https://developers.google.com/nearby/connections/overview, 2019.

\bibitem[He19]{He_2019}
He, Yanzhang; Sainath, Tara~N.; Prabhavalkar, Rohit; McGraw, Ian; Alvarez,
  Raziel; Zhao, Ding; Rybach, David; Kannan, Anjuli; Wu, Yonghui; Pang,
  Ruoming; et~al.: Streaming End-to-end Speech Recognition for Mobile Devices.
\newblock IEEE International Conference on Acoustics, Speech and Signal
  Processing (ICASSP), May 2019.

\bibitem[IS11]{iso-24745}
ISO: , ISO/IEC 24745:2011 Information technology — Security techniques —
  Biometric information protection, 2011.

\bibitem[KJ13]{klare-2012}
{Klare}, B.~F.; {Jain}, A.~K.: Heterogeneous Face Recognition Using Kernel
  Prototype Similarities.
\newblock IEEE Transactions on Pattern Analysis and Machine Intelligence,
  35(6):1410--1422, June 2013.

\bibitem[{Li}07]{stanli-2007}
{Li}, S.~Z.; {Chu}, R.; {Liao}, S.; {Zhang}, L.: Illumination Invariant Face
  Recognition Using Near-Infrared Images.
\newblock IEEE Transactions on Pattern Analysis and Machine Intelligence,
  29(4):627--639, April 2007.

\bibitem[Li17]{liu2017sphereface}
Liu, Weiyang; Wen, Yandong; Yu, Zhiding; Li, Ming; Raj, Bhiksha; Song, Le: ,
  Sphere{F}ace: Deep Hypersphere Embedding for Face Recognition, 2017.

\bibitem[Mo19]{mozilla-dec2019}
Morais, Reuben: , {DeepSpeech} 0.6: {Mozilla’s} Speech-to-Text Engine Gets
  Fast, Lean, and Ubiquitous, 2019.

\bibitem[My19]{mycroft}
Mycroft {AI}, 2019.

\bibitem[Pa16]{gdpr}
Parliament, European~Union: , REGULATION (EU) 2016/679 OF THE EUROPEAN
  PARLIAMENT AND OF THE COUNCIL, 2016.

\bibitem[Pr19]{deepspeech}
Project {DeepSpeech}, 2019.

\bibitem[RU11]{rathgeb2011survey}
Rathgeb, Christian; Uhl, Andreas: A survey on biometric cryptosystems and
  cancelable biometrics.
\newblock EURASIP Journal on Information Security, 2011(1):3, 2011.

\bibitem[Sh19]{shell-gm}
Shell and General Motors Deliver Nationwide In-Dash Fuel Payment.
\newblock
  https://www.shell.us/media/2018-media-releases/shell-and-general-motors-deliver-nationwide-in-dash-fuel-payment.html,
  2019.

\bibitem[SKP15]{facenet}
Schroff, Florian; Kalenichenko, Dmitry; Philbin, James: Face{N}et: A unified
  embedding for face recognition and clustering.
\newblock IEEE Conference on Computer Vision and Pattern Recognition (CVPR),
  Jun 2015.

\bibitem[Sp19]{speechrecognizer}
Speech Recognizer, 2019.

\bibitem[Sq]{square}
Square Terminal.
\newblock https://squareup.com/us/en/hardware/terminal.

\bibitem[St]{rain}
State of In-Car Voice Assistants.
\newblock https://rain.agency/state-of-in-car-voice-assistants/.

\bibitem[SYW13]{nrsr-seltzer2013}
{Seltzer}, M.~L.; {Yu}, D.; {Wang}, Y.: An investigation of deep neural
  networks for noise robust speech recognition.
\newblock In: 2013 IEEE International Conference on Acoustics, Speech and
  Signal Processing.
\newblock pp. 7398--7402, May 2013.

\bibitem[Ta]{visa}
Tap to pay with Visa.
\newblock
  https://usa.visa.com/pay-with-visa/contactless-payments/contactless-payments.html.

\bibitem[TYL17]{pifr-tran2017}
{Tran}, L.; {Yin}, X.; {Liu}, X.: Disentangled Representation Learning GAN for
  Pose-Invariant Face Recognition.
\newblock In: IEEE Conference on Computer Vision and Pattern Recognition
  (CVPR).
\newblock pp. 1283--1292, July 2017.

\bibitem[Wa18]{wang2018cosface}
Wang, Hao; Wang, Yitong; Zhou, Zheng; Ji, Xing; Gong, Dihong; Zhou, Jingchao;
  Li, Zhifeng; Liu, Wei: , Cos{F}ace: Large Margin Cosine Loss for Deep Face
  Recognition, 2018.

\bibitem[We15]{nrsr-enhance2015}
Weninger, Felix; Erdogan, Hakan; Watanabe, Shinji; Vincent, Emmanuel; Le~Roux,
  Jonathan; Hershey, John~R.; Schuller, Bj{\"o}rn: Speech Enhancement with
  {LSTM} Recurrent Neural Networks and its Application to Noise-Robust {ASR}.
\newblock In (Vincent, Emmanuel; Yeredor, Arie; Koldovsk{\'y}, Zbyn{\v{e}}k;
  Tichavsk{\'y}, Petr, eds): Latent Variable Analysis and Signal Separation.
\newblock Springer, Cham, pp. 91--99, 2015.

\bibitem[Wu17]{ranhe-hfr2018}
Wu, Xiang; Song, Lingxiao; He, Ran; Tan, Tieniu: Coupled Deep Learning for
  Heterogeneous Face Recognition.
\newblock CoRR, abs/1704.02450, 2017.

\bibitem[Zh16]{zhang2016joint}
Zhang, Kaipeng; Zhang, Zhanpeng; Li, Zhifeng; Qiao, Yu: Joint face detection
  and alignment using multitask cascaded convolutional networks.
\newblock IEEE Signal Processing Letters, 23(10):1499--1503, 2016.

\end{thebibliography}

\end{document}